\documentclass[twocolumn]{aastex631}
\usepackage{natbib}
\usepackage{amsmath}
\usepackage{appendix}
\usepackage{float}
\makeatletter
\def\@fnsymbol#1{\@arabic{#1}}
\makeatother

\begin{document}
% \nolinenumbers

%\suppressAffiliations
\author[0000-0002-8466-5469]{Collin Cherubim}
\affiliation{Department of Earth and Planetary Sciences, Harvard University, 20 Oxford St., Cambridge, MA 02138, USA}
\affiliation{Center for Astrophysics \textbar \ Harvard \& Smithsonian, 60 Garden St., Cambridge, MA 02138, USA}

\author[0000-0003-1127-8334]{Robin Wordsworth}

\affiliation{School of Engineering and Applied Sciences, Harvard University, 20 Oxford St., Cambridge, MA 02138, USA}

\affiliation{Department of Earth and Planetary Sciences, Harvard University, 20 Oxford St., Cambridge, MA 02138, USA}

\author[0000-0002-0673-4860]{Dan J. Bower}
\affiliation{Institute of Geochemistry and Petrology, Department of Earth and Planetary Sciences, ETH Zurich, Clausiusstrasse 25, Zurich CH-8092, Switzerland}

\author[0000-0002-1462-1882]{Paolo A. Sossi}
\affiliation{Institute of Geochemistry and Petrology, Department of Earth and Planetary Sciences, ETH Zurich, Clausiusstrasse 25, Zurich CH-8092, Switzerland}

\author[0000-0001-9897-9680]{Danica Adams}

\affiliation{Department of Earth and Planetary Sciences, Harvard University, 20 Oxford St., Cambridge, MA 02138, USA}

\author[0000-0003-2215-8485]{Renyu Hu}
\affiliation{Jet Propulsion Laboratory, California Institute of Technology, Pasadena, CA 91109, USA}
\affiliation{Division of Geological and Planetary Sciences, California Institute of Technology, Pasadena, CA 91125, USA}

\correspondingauthor{Collin Cherubim}
\email{ccherubim@g.harvard.edu}

%\title{\textbf{Fractionation of deuterium and helium in sub-Neptune atmospheres: Dependence on atmospheric escape mechanism}}
% \title{The Oxidation Landscape of Sub-Neptunes as Controlled by Atmospheric Escape and Magma Ocean Evolution}

% \title{A Broad View of Small Planet Composition Reveals an Oxidation Gradient Spanning the Radius Valley Sculpted by Atmospheric Escape and Magma Ocean Evolution}

% \title{The Oxidation Gradient Spanning the Small Planet Radius Valley Sculpted by Atmospheric Escape and Magma Ocean Evolution}

% \title{Helium Worlds, Oxygen Worlds, and the Great Oxidation Gradient: A Broad View of Small Planet Evolution Mediated by Atmospheric Escape and Magma Ocean Evolution}

% \title{The Great Oxidation Gradient: A Broad View of Small Planet Evolution Mediated by Atmospheric Escape and Magma Ocean Evolution}

% \title{The Small Planet Oxidation Gradient and the Sub-Neptune to Rocky Planet Transition Mediated by Atmospheric Escape and Magma Ocean Evolution}

\title{An Oxidation Gradient Straddling the Small Planet Radius Valley}

\begin{abstract}
We present a population-level view of volatile gas species (H$_2$, He, H$_2$O, O$_2$, CO, CO$_2$, CH$_4$) distribution during the sub-Neptune to rocky planet transition, revealing in detail the dynamic nature of small planet atmospheric compositions. Our novel model couples the atmospheric escape model {\tt\string IsoFATE} with the magma ocean-atmosphere equilibrium chemistry model {\tt\string Atmodeller} to simulate interior-atmosphere evolution over time for sub-Neptunes around G, K and M stars. Chiefly, our simulations reveal that atmospheric mass fractionation driven by escape and interior-atmosphere exchange conspire to create a distinct oxidation gradient straddling the small-planet radius valley. We discover a key mechanism in shaping the oxidation landscape is the dissolution of water into the molten mantle, which shields oxygen from early escape, buffers the escape rate, and leads to oxidized secondary atmospheres following mantle outgassing. Our simulations reproduce a prominent population of He-rich worlds along the upper edge of the radius valley, revealing that they are stable on shorter timescales than previously predicted. Our simulations also robustly predict a broad population of O$_2$-dominated atmospheres on close-in planets around low mass stars, posing a potential source of false positive biosignature detection and marking a high-priority opportunity for the first-ever atmospheric O$_2$ detection. We motivate future atmospheric characterization surveys by providing a target list of planet candidates predicted to have O$_2$-, He-, and deuterium-rich atmospheres.
\end{abstract}

\section{Introduction} \label{sec:intro}

The NASA Kepler mission revealed that the size of most planets in our galaxy bridges the gap between Earth and Neptune, yet the compositions of these ubiquitous sub-Neptunes remain elusive. Are they gas dwarfs with stratified primordial H/He envelopes, H$_2$O-rich water worlds, scaled-down ice giants, planets with supercritical mixed-layers, Hycean worlds, or something else? Observational surveys seek to answer this question and, so far, it appears to be a varied group, with larger sub-Neptunes hosting hydrogen-dominated atmospheres and smaller sub-Neptunes representing a diverse class of metal-rich planets \citep[e.g.][]{Benneke_2019, Madhusudhan_2023, Piaulet_2024, Benneke_2024, Damiano_2024}.

Hydrodynamic atmospheric escape resulting from X-ray and ultraviolet (XUV)-driven photoevaporation, and potentially core-powered mass loss, is an important aspect of planetary evolution for sub-Neptunes, proposed to have given rise to demographic features such as the small planet radius valley and the hot Neptune desert \citep{Sekiya_1980, Szabo_2011, Beauge_2013, Owen_2013, Jin_2014, Lundkvist_2016, Owen_2017, Fulton_2017, Ginzburg_2018, Vissapragada_2022}. Molecular/atomic diffusion in an escaping atmosphere can lead to chemical stratification and thus mass fractionation, a key predicted aspect of atmospheric evolution with implications for understanding Earth's early atmosphere, which may have been sculpted by escape \citep{Watson_1981, Hunten_1987, Zahnle_1990, Yung_2000, Young_2023}. Signatures consistent with atmospheric escape are also observed on Venus and Mars, each having elevated D/H ratios relative to the proto-solar and terrestrial values, and noble gas anomalies in the case of Mars \citep{Donahue_1982, Kasting_1983, Hunten_1987, Lammer_2020, Mahieux_2024}. Atmospheric fractionation is an important mechanism contributing to the processing of a primary atmosphere and the formation of a secondary atmosphere which is thought to be a central step in the evolution of sub-Neptunes to rocky worlds. Specifically, fractionated atmospheres can become strongly oxidized as they lose hydrogen and retain heavier species, drastically altering planetary chemistry. This process operates regardless of whether the planet has a biosphere and hence can potentially create false positive biosignatures \citep{Domagal-Goldman_2014, Wordsworth_2014, Hu_2015, Luger_2015, Schaefer_2016, Wordsworth_2018}.

Another important aspect of atmospheric and chemical evolution predictions is the thermochemical coupling of the atmosphere with an underlying magma ocean. For planets with Earth-like solid components (i.e. an iron-dominated core and a silicate-dominated mantle in a 1:2 mass ratio), the rocky mantle is expected to be partially molten for surface temperatures $\gtrsim 1500$ K \citep{Hirschmann2000}, depending on the planetary mass; a condition commonly expected for sub-Neptunes with primordial atmospheric mass fractions on the order of tenths of a percent to several percent of the total planetary mass \citep{Lichtenberg_2021}. Magma oceans serve as vast reservoirs for soluble volatile species resulting in planetary interior-atmosphere exchange of key species, especially water, that can later outgas to form secondary atmospheres \citep{Sossi_2020, Lichtenberg_2021, Bower_2022, Gaillard_2022}. Magma ocean formation/evolution in the solar system is thought to be central to the reduction-oxidation state of planetary interiors and atmospheres, and may have set the stage for Earth's long-term atmospheric evolution while also playing a strong role in the bifurcation of climate states between Earth and Venus \citep{ElkinsTanton_2012, Hamano_2013, Armstrong_2019}. Together, atmospheric escape and magma ocean evolution are key controls on the distribution of inherited volatile species in sub-Neptune interiors and atmospheres, which in turn govern planetary chemistry and climate and serve as observable markers of planetary evolution.

Building on \cite{Cherubim_2024}, we set out to map the distribution of key volatiles across the small planet mass-radius-orbital distance landscape by coupling the numerical atmospheric escape model {\tt\string IsoFATE} to the magma ocean-atmosphere equilibrium chemistry model {\tt\string Atmodeller} \citep{Bower_2025}. The speed and flexibility of our model allows for a novel approach to modeling atmospheric escape and outgassing chemistry that explores the full parameter space from sub-Neptunes to airless, rocky planets across a wide range of planetary instellation. This approach enables a population-level view of small planet composition demographics across G, K, and M stars. Hence our model expands the parameter space of previous efforts to model atmospheric fractionation/oxidation via escape, which have focused solely on pure water vapor atmospheres and Earth-analogues
%and have predicted that magma ocean interactions would preclude substantial atmospheric O$_2$ buildup, which our model shows to be incorrect 
\citep{Domagal-Goldman_2014, Wordsworth_2014, Luger_2015, Tian_2015, Schaefer_2016, Bolmont_2017, Wordsworth_2018, Turbet_2020, Johnstone_2020, Krissansen_2024}. Rather than artificially treating enveloped sub-Neptunes and rocky super-Earths as separate planetary regimes---a somewhat illusory dichotomy apart from planets born rocky---our model smoothly captures the transition across the radius valley from enveloped rocky planets to pure rocky planets. In doing so, we show that distinct groups of He-, O$_2$-, CO-, and CO$_2$-dominated planetary atmospheres emerge, as does a gradient of increasing atmosphere/interior oxidation with decreasing planetary radius and orbital distance. Our results serve to inform observational exoplanet surveys, as atmospheric escape/fractionation and magma ocean evolution manifest in observable atmospheric tracers. Importantly, they also provide strong motivation for shifting the focus of atmospheric O$_2$ detection to shorter-period planets around low-mass stars. Our results also have important implications for placing rocky solar system planets in the greater exoplanet context.
%for solar system science because our predictions stem from fundamental atmospheric escape theory used to explain isotopic anomalies on Venus, Earth, and Mars.

We describe the details of our coupled model in Section \ref{sec:model}. We present our key findings in Section \ref{sec:results} and discuss implications of our findings and caveats in Section \ref{sec:discussion}. We discuss how our results inform observational surveys, including a target list, in Section \ref{sec:observations}. Finally, we outline our main conclusions in Section \ref{sec:conclusion}.

\section{Model} \label{sec:model}

\subsection{Atmospheric Escape} \label{sec:escape}

We use the open-source numerical atmospheric escape code {\tt\string IsoFATE}\footnote{IsoFATE source code: \url{https://github.com/cjcollin37/IsoFATE}} \citep{Cherubim_2024}. A key difference between the previous study and the present is that we added C and O so that the model computes molecular diffusion and escape of H (protium), He (helium), D (deuterium), O (oxygen), and C (carbon). {\tt\string IsoFATE} models energy-limited and radiation/recombination-limited XUV-driven photoevaporation, core-powered mass loss, and molecular diffusion to compute variable escape rates for individual species, allowing for mass fractionation of an escaping atmosphere. We model stellar flux evolution as in \cite{Cherubim_2024}. We assume all escaping species are atomic, which represents an upper limit for photolysis efficiency in the upper atmosphere. This assumption is discussed in Section \ref{sec:discussion}.

{\tt\string IsoFATE} simulates atmospheric escape by numerically integrating several differential equations of the general form

\begin{equation}
    \frac{dN_i}{dt} = -A\Phi_i,
    \label{eq:dNdt_main}
\end{equation}

\noindent where $N_i$ is the total number of moles of species $i$, $\Phi_i$ is the number flux of species $i$ [particles m$^{-2}$ s$^{-1}$], and $A$ is the planetary surface area [m$^2$]. All planets in our simulations initially inherit H/He-dominated, solar composition atmospheres. To calculate $\Phi_i$ for the dominant species H and He, we follow the prescription for diffusive fractionation derived in \cite{Wordsworth_2018} \citep[see also][]{Zahnle_1986, Zahnle_1990, Hu_2015, Cherubim_2024}. We calculate all minor species escape fluxes as was done for D by \cite{Cherubim_2024}. Minor species escape fluxes are derived from an analytical expression for an arbitrary number of species escaping in an isothermal, subsonic wind from \cite{Zahnle_1990}:

\begin{equation} \label{eq:Z90}
\begin{split}
& \frac{d\ln f_\mathrm{j}}{dr}  = -\frac{GM_\mathrm{p}(m_\mathrm{j}-m_1)}{kTr^2} - \frac{r_0^2}{r^2}\sum_i[\Phi_\mathrm{i} - f_\mathrm{i}\Phi_1 ] \frac{1}{b_\mathrm{i,1}} \\
& + \frac{r_0^2}{r^2}\sum_i[\Phi_\mathrm{i} - \Phi_\mathrm{j}(f_\mathrm{i} /f_\mathrm{j}) ] \frac{1}{b_\mathrm{i,j}},
\end{split}
\end{equation}

\noindent where $f_\mathrm{j} = N_\mathrm{j}/N_1$ is the mixing ratio of species $j$, $N_1$ is moles of the primary escaping species (i.e., H), $r$ is radius, $r_0$ is the planet radius, $M_\mathrm{p}$ is the planet mass, $m$ is atomic mass, $\Phi$ is escape flux as previously defined, $b_{i,j}$ is the binary diffusion coefficient for species $i$ and $j$, and $G$ is the gravitational constant.

After setting the total number of species to 3, expanding out sums and solving for $\Phi_3$, we arrive at an expression for a minor species (species 3) escape flux:

\begin{equation}
\Phi_3 = f_3 \frac{\Phi_1 + \alpha_3 \Phi_2 + \alpha_2 \Phi_\mathrm{d,2} x_2 - \Phi_\mathrm{d,3}}{1 + \alpha_3 f_2},
\label{eq:Phi3}
\end{equation}

\noindent where $x_i = N_i/N_\mathrm{total}$ (molar concentration), $f_i = x_i/x_1$ (mole fraction), $\alpha_2 \equiv b_\mathrm{H,3}/b_\mathrm{H,He}$, $\alpha_3 \equiv b_\mathrm{H,3}/b_\mathrm{He,3}$ and $\Phi_{\mathrm{d},i} \equiv b_{1,i}/(H_i^{-1} - H_1^{-1})$. We use this formula to calculate the escape flux for all species other than H and He. Equation \ref{eq:Phi3} is valid over all ranges of overall escape flux, having a minimum value of zero.

\subsection{Coupled Atmosphere-Interior Chemistry} \label{sec:magma ocean}

We model magma ocean--atmosphere volatile exchange and equilibrium chemistry with a newly developed Python tool kit for computing the equilibrium conditions at the melt-atmosphere interface called {\tt\string Atmodeller} \citep{Bower_2025}. Given a set of planetary parameters (e.g., surface temperature, planetary mass, radius, mantle melt fraction) and an initial volatile budget, {\tt\string Atmodeller} uses experimentally calibrated solubility laws, together with free energy data for gas species \citep{MZG02}, to determine how volatiles partition between the atmosphere and interior of the planet, ignoring dissolution of volatiles into possible Fe-Ni core phases. We coupled {\tt\string IsoFATE} elemental abundances to {\tt\string Atmodeller} and achieve mass conservation between the two modules in an open system where mass is lost to space and exchanged between the planetary interior and atmosphere. We include H$_2$, He, H$_2$O, O$_2$, CO$_2$, CO, CH$_4$, and deuterium (D), which is treated as chemically equivalent to H, in our model. We specify the mass of the solar composition atmosphere as an initial condition for each planetary simulation, which sets the abundance of each chemical element in our model (H, D, He, C, and O). Then, for a given temperature and mass of H-He-C-O, the partial pressures of volatile gas species at chemical equilibrium are computed following the ideal gas law. These species are allowed to partition between the atmosphere and molten mantle according to solubility laws that are experimentally calibrated for a basaltic composition, since these are most available in the Earth science literature. This yields the converged value of $N_i$, elemental species moles, at any given time step in the simulation. Oxygen fugacity, $f$O$_2$, equivalent to the partial pressure of the dioxygen species in our model, is also calculated at each step by {\tt\string Atmodeller} using the same procedure, that is, based on the pressure, temperature and solubility-modulated elemental composition of the gas phase.

At each time step in our coupled model, we compute the mantle melt fraction as a function of planetary mass and surface temperature following \cite{Wordsworth_2018}. The mantle melt fraction calculation uses a second-order Birch-Murnaghan equation of state to determine the interior structure for an Earth-like planet with a silicate mantle, iron core, and core mass fraction of 0.3, and assumes dry adiabatic convection. We calculate the local mantle melt fraction in the planetary interior as

\begin{equation}
    \psi(r) = 
    \begin{cases}
    0 &: T \leq T_\mathrm{sol} \\
    \frac{T - T_\mathrm{sol}}{T_\mathrm{liq} - T_\mathrm{sol}} &: T_\mathrm{sol} < T < T_\mathrm{liq} \\
    1 &: T \geq T_\mathrm{liq},
    \end{cases}
\label{eq:psi}
\end{equation}

\noindent where $T_\mathrm{sol}$ and $T_\mathrm{liq}$ are the solidus and liquidus temperature, respectively. The total mantle melt fraction is calculated as a function of surface temperature by numerically integrating Equation \ref{eq:psi} in $r$ from the core-mantle boundary to the surface \citep{Wordsworth_2018}. Surface temperature is calculated assuming hydrostatic equilibrium for ideal gases and a dry adiabat for a convective, hydrogen-dominated layer which transitions to an isothermal layer at 0.2 bar \citep{Robinson_2012}. The surface temperature is highly sensitive to the assumed adiabatic index (7/5), which in turn strongly affects the mantle melt fraction. Further, neglecting moist adiabatic effects in our magma oceans leads to slight underestimation of mantle melt fraction. To address these sources of error, we performed sensitivity tests with regard to mantle melt fraction and discuss the results, as well as the potential impact of non-adiabatic temperature profiles on our results, in Section \ref{sec:discussion}.

Our model assumes an upper limit on magma ocean degassing and ingassing efficiency of volatiles because \texttt{Atmodeller} instantaneously re-equilibrates the interior and atmospheric reservoirs when called. This assumption has been often made since the earliest magma ocean models, e.g. \cite{Elkins_2008}, who estimate a complete magma ocean circulation time of 1-3 weeks for terrestrial magma oceans. 
%For degassing to occur, dissolved volatiles must be transported to shallow depths through convection where they exsolve into bubbles which then burst at the surface. Degassing efficiency is therefore mainly dependent on convective velocity \citep{Salvador_2023}. 
In the least efficient case, \cite{Salvador_2023} determine that 0.1 terrestrial oceans of water could take $\approx$ 45,000 yr to degas from a fully molten Earth-mass planet. This timescale is of the same order as a time step in our simulations, i.e., 50,000 yr. For reference, an Earth-mass planet with a 1\% solar-composition atmosphere by mass contains $\approx 0.3$ Earth oceans of atomic oxygen, representing an upper limit on how much water can form. \texttt{Atmodeller} is called every 100 time steps, so all supersaturated water would easily degas for such a planet between each call.

\subsection{Model Simulations} \label{sec:simulations}

We ran a suite of interior/atmosphere evolution models via Monte Carlo simulations over a broad parameter space. For each trial, 500,000 samples were randomly drawn from log uniform grids of initial planet mass $M_\mathrm{p}$ between 1 - 20 M$_\oplus$, initial atmospheric mass fraction $f_\mathrm{atm}$ between 0.1 - 30\%, and orbital period $P$ between 1 - 300 days. The chosen upper limit for $M_\mathrm{p}$ is motivated by the predicted threshold for runaway gas accretion and the chosen range of $f_\mathrm{atm}$ was motivated by feasible values for our planets calculated with gas accretion models \citep{Ginzburg_2016}. Atmospheric mass loss was initiated at 1 Myr. Orbital migration effects were ignored so $P$ was held constant. Model simulations were halted at 5 Gyr---representing the average planetary age in the Milky Way---for the main results, and 10 Gyr in some cases discussed in Section \ref{sec:results}. We assume proto-solar relative abundances of H, He, D, O, and C in a planetary atmosphere enveloping an Earth-like rocky core to simulate nebular gas capture \citep{Lodders_2003}, though in reality these values are expected to deviate for different systems. As such, we employ a ``dry start'' model, ignoring C, H, and O sourced from the planetary interior and sources of volatiles that enhance planetary metallicity such as cometary delivery and planetary formation in ice-rich environments beyond snowlines. Hence, our model does not produce water worlds with water mass fractions on the order of several percent. We simulate escape with XUV-driven photoevaporation alone and the combination of photoevaporation and core-powered mass loss. We ignore scenarios in which core-powered mass loss operates alone, as we expect photoevaporation to dominate in our parameter space of interest \citep{Owen_2023, Tang_2024}. We performed simulations for planets around G, K, and M stars and we focus on the results for planets around M stars given that they are most amenable to atmospheric characterization.

\begin{figure*}
    \centering
    \includegraphics[width=\textwidth]{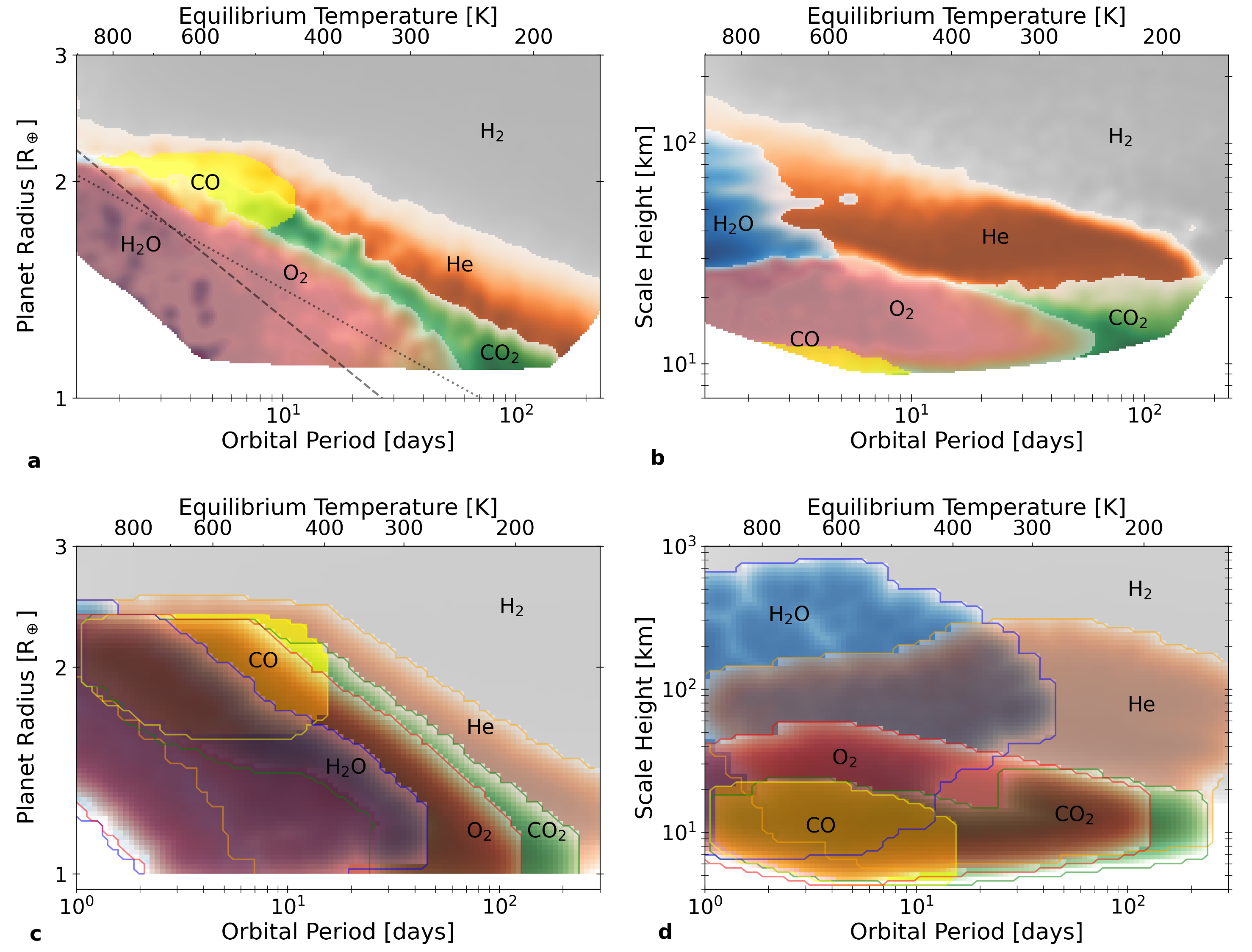}
    \caption{Atmospheric composition trends for planets around M stars after 5 Gyr. Colors are chosen arbitrarily and color intensity indicates species molar concentrations (panels a, b) and simulated planet density (panels c, d) as a function of orbital period vs. planet radius/atmospheric scale height. For panels c and d, each group corresponds to atmospheres with $x_i \geq 50\%$, except H$_2$O, for which the cutoff is 5\%. Contour lines in panels c and d trace the various planet families. Planets trend toward greater oxidation with smaller radii and shorter orbital periods as a result of atmospheric escape-driven fractionation and magma ocean volatile exchange. H$_2$O-rich planets are exceptional in that they are not well confined in $P-R_\mathrm{p}$ space and typically do not reach atmospheric molar concentrations above $\approx$ 10\% as a result of our dry start assumption. The dashed line in panel a shows the radius valley calculated from the present simulations and the dotted line shows that from \cite{Cherubim_2024}. An animated version of this plot is available at: \url{https://github.com/cjcollin37/IsoFATE/blob/main/animated_figures.md}.}
    \label{fig:composition_plot_v2}
\end{figure*}

\begin{figure*}
    \centering
    \includegraphics[width=\textwidth]{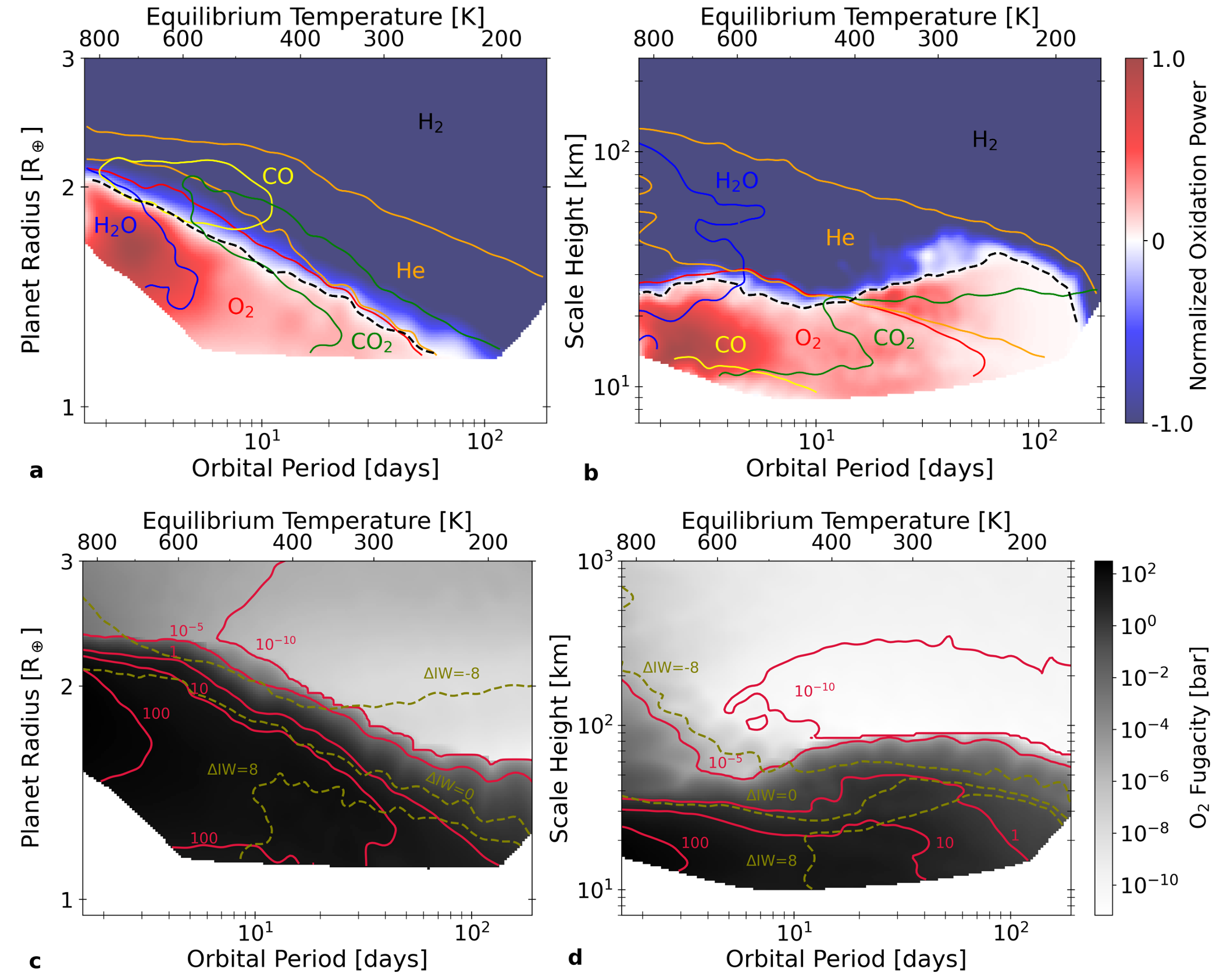}
    \caption{The small planet oxidation gradient around M stars after 5 Gyr. Contour lines in panels a and b represent boundaries for atmospheric compositions shown in Figure \ref{fig:composition_plot_v2}. The colors in panels a and b show normalized oxidation power for all volatiles in the interior and atmosphere. This metric quantifies the number of electrons each atomic species can accept in a redox reaction. The red region indicates a net oxidizing planet in which the oxidizing species O dominates over reducing species. The grayscale in panels c and d shows O$_2$ fugacity, $f$O$_2$, in bar. The red contours outline the boundaries at specified $f$O$_2$ values and the green dashed contours outline boundaries for $f$O$_2$ relative to the iron-wüstite mineral buffer evaluated at 1 bar: $\Delta$IW = $log_{10}f$O$_2 - log_{10}$IW.}
    \label{fig:oxidation_plot_v2}
\end{figure*}

\section{Results} \label{sec:results}

\subsection{Radius Valley Redox Gradient} \label{sec:redox}

\begin{figure*}
    \centering
    \includegraphics[width=\textwidth]{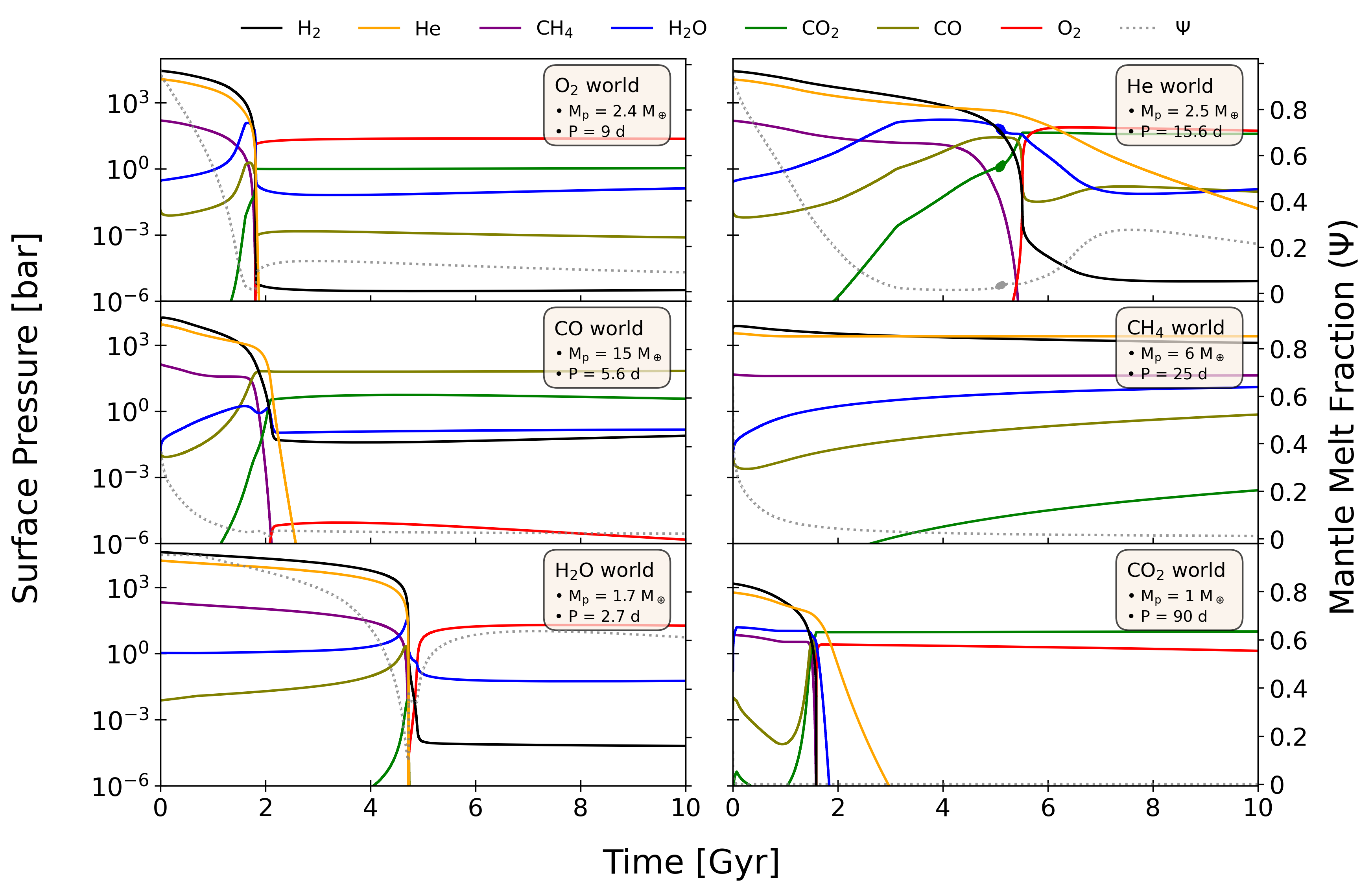}
    \caption{Atmospheric composition evolution over time of archetypical planets from each of the families shown in Figure \ref{fig:composition_plot_v2}. Species partial pressures are represented by the colored lines, corresponding to the left axes and mantle melt fraction, $\Psi$, is represented by the gray dotted line, corresponding to the right axes. Each archetype is defined based on its composition at 5 Gyr, but the longer term fate of each archetype is shown to 10 Gyr.}
    \label{fig:archetype_plot}
\end{figure*}

Our Monte Carlo simulations reveal an atmospheric oxidation gradient that increases toward shorter orbital periods and smaller planet radii (Figures \ref{fig:composition_plot_v2} and \ref{fig:oxidation_plot_v2}). Figures \ref{fig:composition_plot_v2}a and \ref{fig:composition_plot_v2}b show interpolated species molar concentrations as a function of orbital period vs. planet radius and scale height, $H = k_\mathrm{B} T_\mathrm{eq}/\mu g$, where $k_\mathrm{B}$ is the Boltzmann constant, $T_\mathrm{eq}$ is the planet equilibrium temperature, $\mu$ is the average atmospheric atomic mass, and $g$ is the gravitational field strength. Figures \ref{fig:composition_plot_v2}c and \ref{fig:composition_plot_v2}d show the density of simulated planets with atmospheres dominated by each indicated species, except for H$_2$O, for which the threshold is 5\%. 
%Simulated planets were placed on regular $P-R_\mathrm{p}$ and $P-H$ grids...

Five broad families of atmospheres with varying oxidation states emerge in the planet radius-orbital period parameter space that determine the dominant atmospheric species: H$_2$ worlds, He worlds, CO worlds, CO$_2$ worlds, and O$_2$ worlds. We exclude H$_2$O worlds from this list as our simulations produce few planets ($< 1\%$) with water-dominated atmospheres, as discussed in Section \ref{sec:discussion}. H$_2$-dominated worlds represent the most reduced planets, having retained H/He-dominated primordial atmospheres. He worlds have typically lost the bulk of their H, making them depleted in H$_2$ and typically enriched in CO and CO$_2$, or, less commonly, CH$_4$ if they retain sufficient H$_2$. CO worlds are more oxidized than He worlds. They are H$_2$ poor, CO$_2$ rich, and typically have short orbital periods, existing within and just above the radius valley. CO$_2$ worlds are even more oxidized. Their second most abundant atmospheric species tends to be O$_2$ and they have smaller radii than the He and CO worlds, situated between the He worlds and O$_2$ worlds. Finally, O$_2$ worlds represent the most oxidized planets, possessing O$_2$-dominated atmospheres. They exist within and below the radius valley, have the smallest radii, and are on close-in orbits. H$_2$O-rich worlds are rarely H$_2$O-dominated in our simulations and span a wide parameter space that largely overlaps with the other families, particularly O$_2$ worlds.

It is important to note that overlapping families in Figure \ref{fig:composition_plot_v2} do not necessarily indicate the prevalence of individual planets belonging to multiple families. For example, the CO- and O$_2$ world groups overlap substantially in panels c and d, but atmospheres cannot possess both CO and O$_2$ in high abundances. O$_2$-dominated atmospheres are typically accompanied by CO$_2$ as the second most abundant gas, and vice versa. The CO and O$_2$ world groups overlap in $P-R_\mathrm{p}$ space because CO-dominated worlds can exist where O$_2$-dominated worlds can independently exist. Interestingly, CO worlds must start with much smaller atmospheres ($f_\mathrm{atm,0} \approx 0.005$) compared with O$_2$ worlds ($f_\mathrm{atm,0} \approx 0.02$). Both end up with similar atmospheric masses after 5 Gyr, and since CO and O$_2$ have similar molecular masses, their radii are similar for a given planetary mass, resulting in the observed overlap in $P-R_\mathrm{p}$ space. This implies that we may infer formation conditions based on a planet's atmospheric oxidation state, mass, and radius. Potential photochemistry effects are discussed in Section \ref{sec:photochemistry}.

Figure \ref{fig:oxidation_plot_v2} a and b show interpolated normalized oxidation power, $P_\mathrm{ox}$, with overplotted contours of the same families presented in Figure \ref{fig:composition_plot_v2}. Oxidation power, $P_\mathrm{ox}$, is determined following \cite{Wordsworth_2018}. Each elemental species is assigned an oxidation potential, $p_i$, which essentially reflects the number of electrons available to exchange in a reduction-oxidation reaction: $p_\mathrm{O} = +2$, $p_\mathrm{H} = -1$, and $p_\mathrm{C} = -4$. For each simulated planet, $p_\mathrm{i}$ is summed for all species to get the total oxidation power, $P_\mathrm{ox} = \sum_{i} n_i p_i$, where $n_i$ is the total number of atoms of species $i$. Figure \ref{fig:oxidation_plot_v2} a and b show the interpolated $P_\mathrm{ox}$ values normalized between -1 and 1. Only planets in the red parameter space have sufficient O to be net oxidized and hence exist within the O$_2$ world family.

Figure \ref{fig:oxidation_plot_v2} c and d show the O$_2$ fugacity, $f$O$_2$, for the same simulated planets plotted in panels a and b. The solid red contours show $f$O$_2$ values in bar and the dashed green contours indicate the logarithmic ratio between the $f$O$_2$ (in bar) and that defined by the iron-wüstite (IW) mineralogical buffer: $\Delta$IW = $log_{10}f$O$_2 - log_{10}$IW \citep{Hirschmann2021}. Planets with small radii and lower equilibrium temperatures have greater $\Delta$IW values than small, hotter planets on closer-in orbits. This is because the $f$O$_2$ in equilibrium with the iron-wüstite buffer increases with temperature. Therefore, at constant $\Delta$IW, higher temperatures result in higher $f$O$_2$ (in bar). We find a population of planets with lower surface temperatures (mean $\approx 1840$ K) that are \textit{more} oxidized, relative to IW (above $\Delta$IW = +8), than are those with higher surface temperatures (mean $\approx 3000$ K) and higher absolute $f$O$_2$ ($>$100~bar, but between IW and $\Delta$IW = +8). %have different temperature dependencies, hence why the most oxidized simulated planets occupy a different region of parameter space than those with the greatest $\Delta$IW values. 
For reference, the present-day Earth's mantle has $\Delta$IW = +3.5 at 1650 K \citep{frostmccammon2008}.
%%% note how different these planets are from Earth? deltaIW = 0 planets have much higher fO2 than Earth due to higher surface temp I think.

\begin{figure*}
    \centering
    \includegraphics[width=\textwidth]{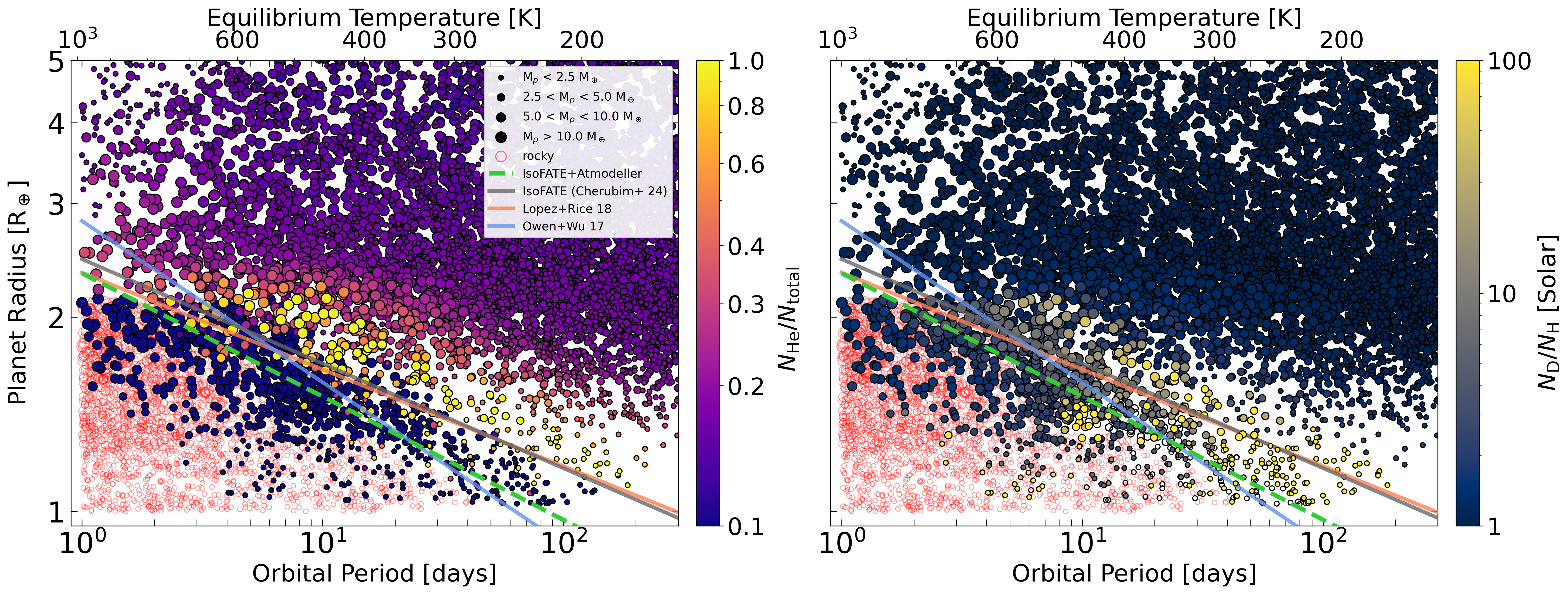}
    \caption{He concentration (left) and D/H ratio (right) for simulated planets around an M1 type star after 5 Gyr. Allowing interior exchange with the molten mantle results in a large population of planets with thin, high mean molecular weight atmospheres in and below the radius valley whose compositions are shown in Figures \ref{fig:composition_plot_v2} and \ref{fig:oxidation_plot_v2}. An animated version of this plot is available at: \url{https://github.com/cjcollin37/IsoFATE/blob/main/animated_figures.md}.}
    \label{fig:He_D_plot}
\end{figure*}

The atmospheric composition evolution of six archetypical worlds orbiting a typical M1 type star from each oxidation family is shown in Figure \ref{fig:archetype_plot}. The archetypes can be categorized into three general groups illustrated in Figure \ref{fig:archetype_plot}: planets with atmospheres in which the dominant gas is H$_2$ (as for the CH$_4$ world and the H$_2$ worlds, not shown), O$_2$/CO$_2$ (as for the O$_2$, CO$_2$, H$_2$O, and He worlds), or CO/CO$_2$ (as for the CO world). The second most abundant species in O$_2$- and CO world atmospheres is CO$_2$ and the second most abundant species in CO$_2$ atmospheres is O$_2$. H$_2$O can be abundant or depleted across all groups, but tends to be retained at least at the ppm level across groups. No planets in our simulations form CH$_4$-dominated atmospheres, although some atmospheres can reach CH$_4$ molar concentrations up to several percent. The extended 10 Gyr simulations reveal that He worlds are often transient phenomena (discussed more in Section \ref{sec:helium}) and tend to be rich in oxidized species such as O$_2$, CO, and CO$_2$. Figure \ref{fig:archetype_plot} illustrates that planet atmospheric composition can change drastically within the first 5 Gyr of evolution, sometimes earlier. The archetypes show that planets are highly dynamic objects with atmospheres and interiors that change composition significantly over Gyr timescales. 

Figure \ref{fig:archetype_plot} also gives a sense of the typical magma ocean lifetimes for planets in our simulations. Most simulated planets maintain non-zero mantle melt fractions throughout the typical 5 Gyr simulation time, although only $\approx$ 50\% of CO$_2$ worlds do as a result of their low atmospheric masses and wide incident stellar flux range. Roughly one third of He worlds have solid mantles by 5 Gyr, as they occupy a similar stellar flux range but tend to have larger atmospheric masses than CO$_2$ worlds, leading to greater surface temperatures. Planets with non-H$_2$-dominated atmospheres with $P \leq 4$ days (semi-major axis $\leq$ 0.04 AU, $T_\mathrm{eq} \geq 620$ K) tend to have mantle melt fractions $\approx$ 20 - 80\%, while those with $P \geq 50$ days (semi-major axis $\geq$ 0.21 AU, $T_\mathrm{eq} \leq 267$ K) tend to have solid mantles. Planets in between have melt fractions between 0 - 20\%.

The most oxidized planets in our simulations reside below the radius valley and have thin, high mean molecular weight atmospheres largely supplied by volatiles that outgas from the mantle as it crystallizes. Such planets are visible in Figure \ref{fig:He_D_plot} and were not produced in our initial {\tt\string IsoFATE} simulations, which did not include coupled atmosphere-interior chemistry \citep{Cherubim_2024}. The dominant physical mechanism giving rise to such planets is water dissolution in the liquid mantle. Water is highly soluble in silicate liquids \citep{Dixon_1995,Sossi_2023}, particularly compared to the other volatile species considered in our system, and the vast majority ($> 99~\%$) of its mass is sequestered in the mantle early in a planet's evolution. This has the effect of modulating the rate of atmospheric escape. When water is dissolved in the mantle, it is shielded from escape, thereby lowering the escape rate because the planetary radius decreases. This effect is magnified for larger atmospheres because they produce higher surface temperatures which increases the volume of the magma ocean. As the primordial atmosphere escapes, water gradually outgasses, is photolyzed, and escapes as H and O. Because H escapes more readily than O, this process must result in oxidation of the atmosphere via Rayleigh distillation, given that the budgets of all volatile elements are finite. 

\begin{figure}
    \centering
    \includegraphics[width=\hsize]{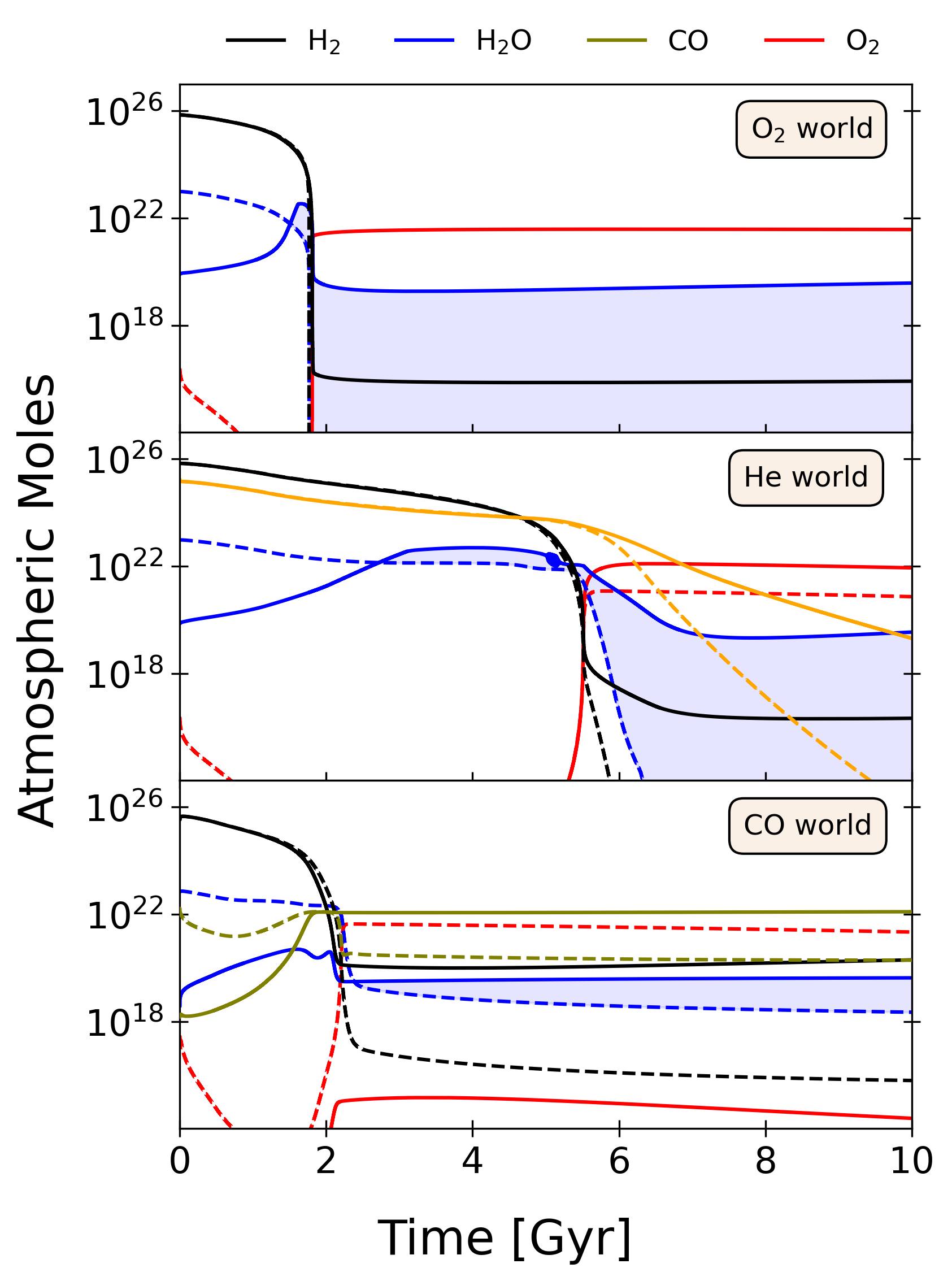}
    \caption{Select archetypes demonstrating the mantle water shielding mechanism. Dashed lines show simulations in which H$_2$O was not allowed to dissolve in the molten mantle and solid lines show standard simulations with H$_2$O dissolution included. The blue shaded region highlights the atmospheric H$_2$O that is retained purely due to H$_2$O dissolution in the mantle. The water shielding mechanism also stabilizes atmospheric H$_2$ on Gyr timescales and enables greater atmospheric mass retention generally.}
    \label{fig:archetype_water_shielding}
\end{figure}

The water shielding mechanism allows planets to retain atmospheres that would otherwise lose them entirely and it allows H$_2$ to be retained on longer timescales (several Gyr). The mechanism and its impact on three archetypical worlds is shown in Figure \ref{fig:archetype_water_shielding}. The dashed lines represent simulations in which water solubility in the mantle was excluded and the solid lines represent simulations with solubilities included for all species. The blue shaded regions highlight the amount of water retained as a direct result of the water shielding mechanism enabled by water dissolution in the mantle. For the O$_2$ world, without water shielding, the entire atmosphere would be stripped within 1 Gyr. In the case of the CO world, water shielding is shown to be a crucial mechanism for atmospheric oxidation resulting in CO as the dominant species. In all cases, substantially more H$_2$ is retained at least up to 10 Gyr.

\subsection{Helium and Deuterium Worlds} \label{sec:helium}

\begin{figure}
    \centering
    \includegraphics[width=\hsize]{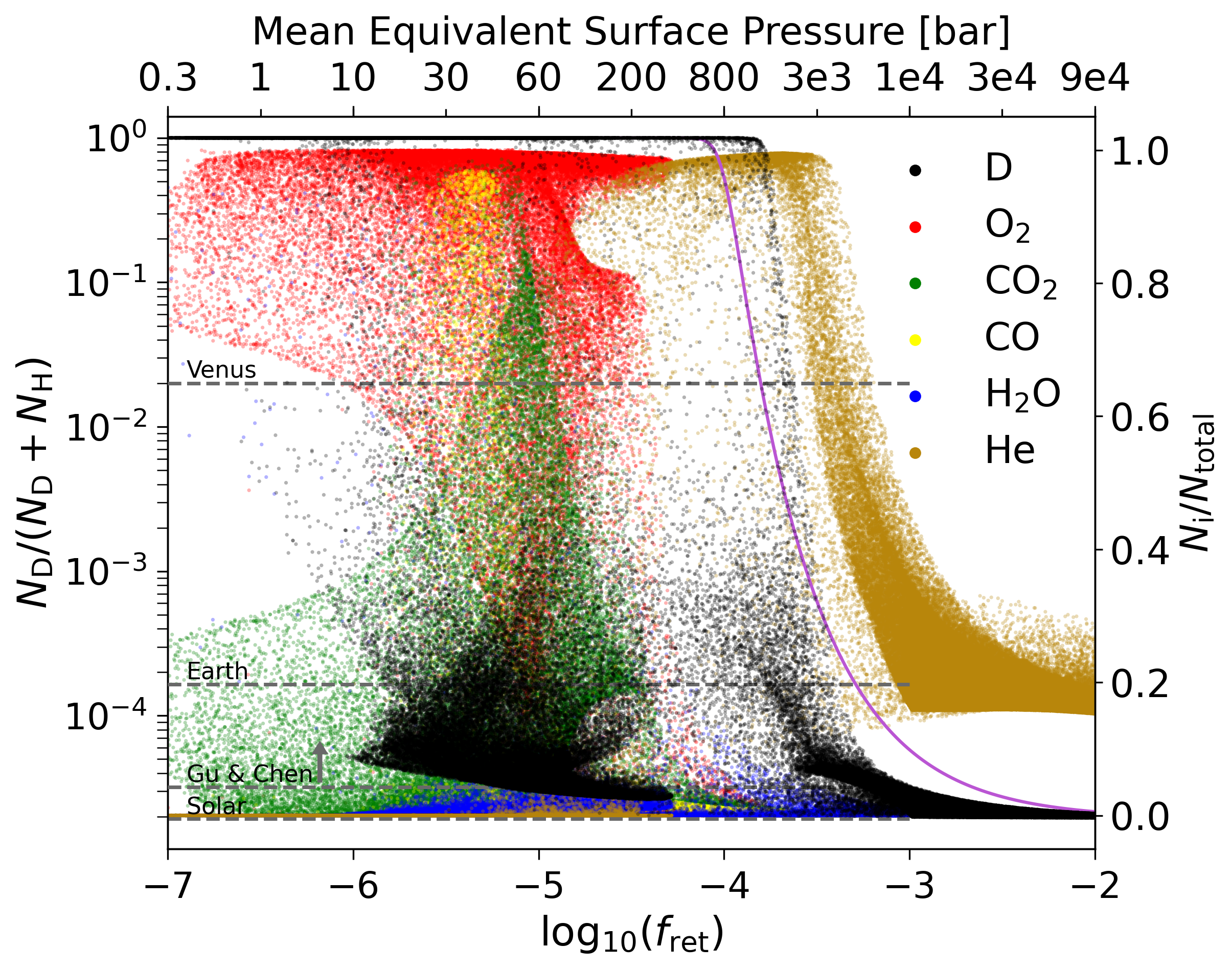}
    \caption{Molar concentration of atmospheric species as a function of retained atmospheric mass fraction $f_\mathrm{ret}$ for simulated planets around an M1 star after 5 Gyr of atmospheric escape and magma ocean evolution. Molar concentration of D is expressed relative to H as black points corresponding to the left axis. All other points correspond to the right axis. The upper axis corresponds to an approximated mean surface pressure for each $f_\mathrm{ret}$ bin. The purple line shows an analytic solution for molar concentration of D as a function of $f_\mathrm{ret}$ and planetary equilibrium temperature $T_\mathrm{eq}$ (see \cite{Cherubim_2024} for derivation). The solution shown here is for a mean $T_\mathrm{eq}$ = 500 K. Venus and Earth dashed lines show D concentrations corresponding to Venusian D/H and the Vienna Standard Mean Ocean Water (VSMOW) values respectively. The ``Gu \& Chen'' dashed line shows the maximum D/H value reported by \cite{Gu_2023} in their photoevaporation simulations.}
    \label{fig:DH_plot}
\end{figure}

Sub-Neptunes with fractionated atmospheres enriched in He \citep{Hu_2015, Malsky_2020, Malsky_2023, Cherubim_2024} and D \citep{Gu_2023, Cherubim_2024} have been robustly predicted by atmospheric escape models. Our {\tt\string IsoFATE}-{\tt\string Atmodeller} coupled model reproduces the same prominent population of He worlds along the upper edge of the radius valley as previously predicted with {\tt\string IsoFATE} in \cite{Cherubim_2024}. The He worlds can be seen in Figures \ref{fig:composition_plot_v2}, \ref{fig:oxidation_plot_v2}, and \ref{fig:He_D_plot}. The strongly D-enriched planet population predicted by \cite{Cherubim_2024} is also reproduced, but the correlation between D/H and atmospheric surface pressure is weaker (Figure \ref{fig:DH_plot}). This is due to the water shielding mechanism described in Section \ref{sec:redox}. Water is the dominant H carrier for a planet that has lost the bulk of its primordial (i.e., H-, He-rich) atmosphere and is highly soluble in the magma ocean. This serves as a vast H reservoir that provides a steady supply of H to the atmosphere, thereby diluting the elevated D/H ratio in the atmosphere.
%This serves as a vast H reservoir that dilutes the elevated D/H ratio that develops in the atmosphere via late-stage outgassing, thereby providing a steady supply of unfractionated (i.e., solar) H to the atmosphere.
As a result, thinner atmospheres are allowed to have lower D/H ratios than previously predicted in {\tt\string IsoFATE} simulations without magma ocean coupling. Our simulations suggest that D/H may still be used to place an upper limit on atmospheric mass and surface pressure. While most of our D-enriched planets peak near Earth's value of $\approx$ 10 $\times$ solar, many planets exceed Venus-like values and can even be depleted entirely of H in place of D. Planets with enhanced D/H tend to have $M_p <$ 5 M$_\oplus$, averaging $\approx 2$ M$_\oplus$; orbital periods between 1 - 150 days, averaging 20-30 days; and initial $f_\mathrm{atm} \leq 1\%$.

While the He world population is robustly predicted and dominates much of the parameter space, Figure \ref{fig:archetype_plot} demonstrates that they are actually transient states that typically exist on the order of hundreds of Myr to several Gyr. The archetypes in Figure \ref{fig:archetype_plot} were selected based on their atmospheric compositions at 5 Gyr, which we take to be a representative age for an average exoplanet around a main sequence star in the Milky Way. However, when extending the simulations to 10 Gyr, we see that He is not stable on longer timescales and is gradually lost. If the He inventory manages to survive the bulk EUV emission (within 500 Myr for M stars in our model), a He-dominated atmosphere can remain stable for several Gyr. However, even as EUV emission rapidly declines after the stellar saturation phase, the energy received by the planet is sufficient to drive off He. Unlike H$_2$O, which can be sequestered in the mantle during the magma ocean phase, and gradually released as the mantle crystallizes, He is inert and is relatively insoluble in magma, leaving it susceptible to escape \citep{Jambon1986}. The He world survival timescale is therefore sensitive to assumptions about planetary mass, surface temperature, and stellar evolution. For simulated planets around M stars whose atmospheres become He-dominated at some point in their lifetimes, the mean survival timescale is 900 Myr and the median is 380 Myr.

An additional factor that could contribute to prolonging a He world's lifetime is radiogenic He produced by uranium and thorium decay in the mantle. Extrapolating from the estimated He outgassing rate from Earth's surface today, we expect $\approx 1 \times 10^{18} - 1 \times 10^{21}$ mol/Gyr for rocky planets in the 1 - 10 M$_\oplus$ range \citep{Krasnopolsky_1994}. For common He escape rates, e.g. $\approx 1\mathrm{x}10^{23}$ mol/Gyr for the archetypical He world, radiogenic He may be insufficient to replenish escaping He in most cases, however the Earth estimate may represent a lower bound since He outgassing on planets with magma oceans is not expected to be limited by tectonic activity as it is on Earth.

\subsection{Oxygen Worlds} \label{sec:oxygen}

\begin{figure}
    \centering
    \includegraphics[width=\hsize]{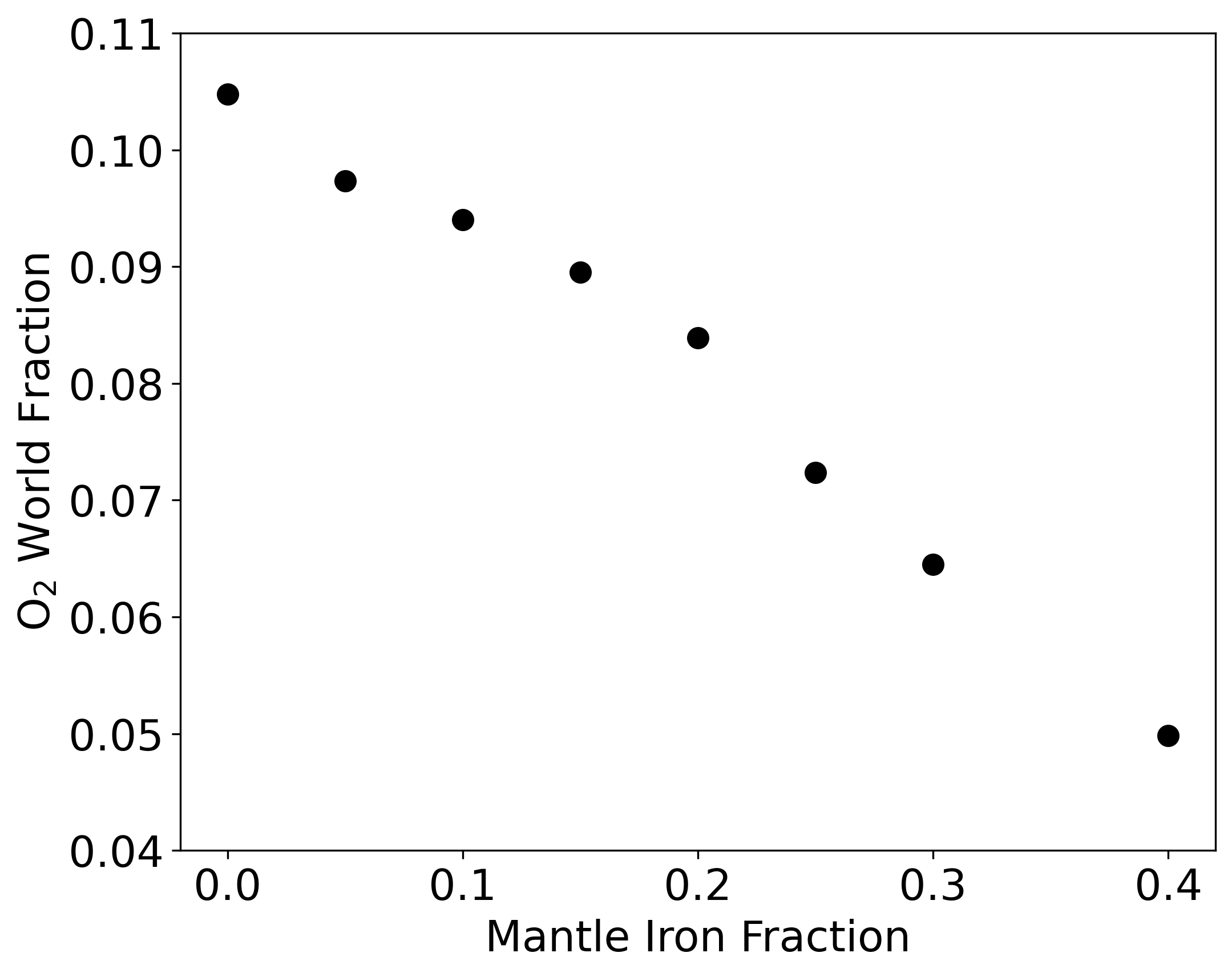}
    \caption{Fraction of all simulated planets around M stars with atmospheres that have O$_2$-dominated atmospheres as a function of assumed mantle iron mass fraction.}
    \label{fig:Fe_plot}
\end{figure}

The prominent O$_2$ world population seen in Figures \ref{fig:composition_plot_v2} and \ref{fig:oxidation_plot_v2} comprises $\approx 70 \%$ of all simulated planets with non-H$_2$ secondary atmospheres. This represents an upper limit on atmospheric oxidation and O$_2$ enrichment because our magma ocean model ignores iron oxidation reactions. That is, if the magma ocean is communicating with the atmosphere throughout a planet's lifetime, then the presence of Fe dissolved as Fe$^{2+}$ and Fe$^{3+}$ in the mantle of the planet could buffer the O$_2$ levels in the atmosphere to lower proportions than those shown in Figures \ref{fig:composition_plot_v2} and \ref{fig:oxidation_plot_v2}.
%That is, reactions of the type  FeO(l,s) + H$_2$(g) = Fe(l,s) + H$_2$O(g) \citep[e.g.,][]{Kite_2020} are neglected, which would otherwise act to maintain a constant O$_2$ fugacity, \textit{f}O$_2$, in the atmosphere, via the reaction FeO(l,s) = Fe(l,s) + 1/2O$_2$(g), provided the moles of Fe, $N_\mathrm{Fe} >> N_\mathrm{H}$. 
We performed simulations to test the sensitivity of O$_2$ world occurrence to assumptions about iron abundance and oxidation state in the mantle, following the approach in \cite{Schaefer_2016} and \cite{Wordsworth_2018}. We assumed mantle iron mass fractions between 0.05 - 0.4 and that all iron was initially in a reduced state: Fe$^{2+}$. At each time step of our numerical simulation, any atmospheric O$_2$ buildup was removed to oxidize Fe$^{2+}$ to Fe$^{3+}$ via 4FeO + O$_2$ $\rightarrow$ 2Fe$_2$O$_3$. This reaction proceeded until all Fe$^{2+}$ was oxidized to Fe$^{3+}$, at which point O$_2$ was allowed to accumulate in the atmosphere. The O$_2$ world population is recovered in these simulations and is robust to the mantle iron sink, as seen in Figure \ref{fig:Fe_plot}. O$_2$ world occurrence decreases with increasing mantle iron mass fraction, as O in the atmosphere is consumed through the oxidation of ferrous iron, but is only marginally affected by mantle iron mass fractions estimated for the rocky solar system bodies, e.g. $\approx 6 \%$ for Earth \citep{McDonough_1995}. That the occurrence of O$_2$ worlds is largely insensitive to the presence or absence of a mantle iron sink, despite Fe being more abundant than O, revealed that the vast majority of such worlds tend to have smaller and/or more short-lived magma oceans, preventing all of the atmospheric O$_2$ from being consumed by the reduced iron reservoir in the mantle. In other words, mantle freezing outpaces O$_2$ production sufficiently to preclude substantial O$_2$ sequestration via mantle iron oxidation.

A competing factor to atmospheric O$_2$ buffering due to mantle iron oxidation is enhanced planetary metallicity. Our simulations generate sub-Neptunes that possess C, H, and O in solar abundances and hence are initially ``dry,'' though this ignores O in the interior, which could combine with the H delivered in the atmosphere to form H$_2$O (see above). For planets that inherit additional water, for example through cometary delivery or accretion of icy planetesimals during formation, the O inventory can be enhanced by several orders of magnitude. If the O abundance is sufficiently enhanced on wet planets endowed with larger water inventories, O$_2$ may continue to build up in the atmosphere as mantle iron oxidation saturates, H$_2$O is photolyzed, and H escapes. \cite{Tian_2015} showed that atmospheric O$_2$ buildup correlates with initial water inventory for Earth analogues around early M stars undergoing atmospheric escape.

%%% add additional simulation results here.

\section{Observational Prospects} \label{sec:observations}

\begin{deluxetable*}{lcccccccccc}
\tabletypesize{\small}
\tablecaption{Fractionated planet candidates predicted from $M_\mathrm{p}$-$R_\mathrm{p}$-$F_\mathrm{p}$ interpolation \label{tab:targets}}
%\tablewidth{0pt}
\tablehead{\colhead{Planet} & \colhead{Stellar type} & \colhead{M$_\mathrm{p}$ [M$_\oplus$]} 
& \colhead{R$_\mathrm{p}$ [R$_\oplus$]} & \colhead{P [days]} & \colhead{T$_\mathrm{eq}$ [K]} & \colhead{J-band mag} & \colhead{TSM} & \colhead{$x_\mathrm{He} > 0.3$} & \colhead{D/H $> 10$} & \colhead{$x_\mathrm{O2} > 0.2$}}
\startdata
\\ %% M STARS
TRAPPIST-1 f & M & 1.039 & 1.045 & 9.20754 & 218 & 11.354 & 113 & 0.01 & 0.97$^{*}$ & 0.94 \\
TRAPPIST-1 c & M & 1.308 & 1.097 & 2.421937 & 341 & 11.354 & 161 & 0.03 & 0.89$^{*}$ & 0.94 \\
TRAPPIST-1 g & M & 1.321 & 1.129 & 12.352446 & 198 & 11.354 & 101 & 1.0 & 1.0$^{*}$ & 0.05 \\
TRAPPIST-1 b & M & 1.374 & 1.116 & 1.510826 & 399 & 11.354 & 189 & 0.05 & 0.73$^{*}$ & 0.94 \\
LTT 1445 A c & M & 1.54 & 1.147 & 3.1239035 & 513 & 7.294 & 302 & 0 & 0.60$^{*}$ & 0.51 \\
GJ 1132 b & M & 1.84 & 1.192 & 1.62892911 & 583 & 9.245 & 198 & 0 & 0.54$^{*}$ & 0.69 \\
GJ 357 b & M & 1.84 & 1.2 & 3.9306 & 519 & 7.337 & 181 & 0 & 0.54$^{*}$ & 0.53 \\
LHS 1140 c & M & 1.91 & 1.272 & 3.77794 & 426 & 9.612 & 143 & 1 & 0.9 & 0.45$^{**}$ \\
L 98-59 d & M & 1.94 & 1.521 & 7.4507245 & 409 & 7.933 & 269 & 0.35 & 0 & 0 \\
Kepler-138 d & M & 2.1 & 1.51 & 23.0923 & 379 & 10.293 & 24 & 0.4 & 0 & 0 \\
HD 260655 b & M & 2.14 & 1.24 & 2.76953 & 710 & 6.674 & 191 & 0 & 0 & 0.68 \\
Kepler-138 c & M & 2.3 & 1.51 & 13.7815 & 450 & 10.293 & 25 & 0.52 & 0 & 0 \\
LHS 1478 b & M & 2.33 & 1.242 & 1.9495378 & 600 & 9.615 & 119 & 0 & 0.36 & 0.53 \\
K2-3 c & M & 2.68 & 1.582 & 24.646729 & 373 & 9.421 & 30 & 0.31 & 0 & 0 \\
TOI-244 b & M & 2.68 & 1.52 & 7.397225 & 459 & 8.827 & 70 & 0.64 & 0 & 0 \\
TOI-1452 b & M & 4.82 & 1.672 & 11.06201 & 329 & 10.604 & 39 & 0.65 & 0 & 0 \\
TOI-776 b & M & 5 & 1.798 & 8.24662 & 521 & 8.483 & 51 & 0.24 & 0 & 0 \\
LHS 1140 b & M & 5.6 & 1.73 & 24.73723 & 228 & 9.612 & 66 & 0.59 & 0 & 0 \\
TOI-1695 b & M & 6.36 & 1.9 & 3.1342791 & 701 & 9.64 & 42 & 0.15 & 0 & 0 \\
\\ %% K STARS
TOI-260 b & K & 3.3 & 1.473 & 13.475815 & 480 & 7.376 & 65 & 0 & 0.25 & 0 \\
K2-36 b & K & 3.9 & 1.43 & 1.422614 & 1330 & 10.034 & 24 & 0 & 0.02$^{*}$ & 0 \\
HIP 29442 d & K & 5.14 & 1.538 & 6.429575 & 973 & 8.056 & 22 & 0 & 0.06 & 0.42 \\
GJ 9827 b & K & 5.14 & 1.577 & 1.2089819 & 1173 & 7.984 & 79 & 0 & 0 & 0.26 \\
TOI-1235 b & K & 6.69 & 1.69 & 3.444714 & 776 & 8.711 & 33 & 0 & 0.29 & 0 \\
HD 15337 b & K & 7.2 & 1.699 & 4.75642 & 990 & 7.553 & 37 & 0 & 0 & 0.28 \\
\\ %% G STARS
K2-265 b & G & 7.3 & 1.708 & 2.36902 & 1383 & 9.726 & 16 & 0 & 0 & 0.22 \\
TOI-1451 b & G & 15.2 & 2.611 & 16.537944 & 793 & 8.38 & 24 & 0.01 & 0 & 0 \\
Kepler-131 b & G & 16.13 & 2.41 & 16.092 & 778 & 10.418 & 7 & 0.09 & 0 & 0 \\
TOI-1753 b & G & 16.6 & 2.479 & 5.3846104 & 1099 & 10.553 & 10 & 0.02 & 0 & 0 \\ 
TOI-1751 b & G & 19.5 & 2.951 & 37.46852 & 711 & 8.251 & 15 & 0.01 & 0 & 0 \\
\enddata
\tablecomments{Planet parameters were obtained from the \dataset[NASA Exoplanet Archive]{\doi{10.26133/NEA13}}, queried on October 7, 2024. Only planets with $R_\mathrm{p}$ and $M_\mathrm{p}$ precision $\geq$ 10 \% and 50 \%, respectively, were included. TSM calculation follows prescription in \cite{Kempton_2018}. The last three columns show the probability of fractionation based on the number of fractionated planets divided by the total number of planets in each grid cell occupied by the target and its error bars in $M_\mathrm{p}$-$R_\mathrm{p}$-$F_\mathrm{p}$ space. $x_\mathrm{i}$ represents molar concentration. D/H represents molar ratio relative to the solar value. All predictions are based on 5 Gyr EUV-driven photoevaporation simulations.
\newline $^{*}$ D/H $> 100$
\newline $^{**}$ $x_\mathrm{O2} > 0.01$
  %1) \cite{}
  %2) \cite{}
}
\end{deluxetable*}

Short-period, abiotically oxidized planets, like those shown in Figure \ref{fig:composition_plot_v2}, present a unique opportunity to detect O$_2$ atmospheres on rocky planets, owing to their large scale heights, high H$_2$O/O$_2$ abundances, higher transit probability, and more frequent transits. To date, no detection of atmospheric O$_2$ has been made on an exoplanet, and proposed observations with the next-generation Extremely Large Telescopes (ELTs) have exclusively focused on targeting Earth twins in habitable zones \citep{Snellen_2013, Rodler_2014, Domagal-Goldman_2014, Serindag_2019, Lopez-Morales_2019, Hardegree-Ullman_2023}. Current predictions estimate upwards of 20 transits are necessary in the best case scenarios for detecting atmospheric O$_2$ on Earth-like (i.e. similar size and stellar flux) exoplanets orbiting hypothetical nearby low-mass stars, and upwards of 50 transits in less ideal scenarios \citep{Serindag_2019, Lopez-Morales_2019, Hardegree-Ullman_2023}. While Earth analogues are compelling targets, detecting O$_2$ on short-period, abiotically oxidized worlds is a technically less demanding strategy that would require significantly fewer transits to achieve sufficient SNR with the ELTs. Furthermore, theory that predicts abiotic production of O$_2$ on rocky planets is well established and based on a simple physical mechanism. In contrast, biotic O$_2$ on Earth-like exoplanets requires a significantly more complex history, i.e., biogenesis and all its prerequisites, followed by photosynthesis and a Great Oxidation Event \citep[][]{sessions2009continuing,lyons2021oxygenation}. 

Abiotically oxidized worlds are a robust prediction of evolutionary models, while no model yet exists to predict the probability of biogenic O$_2$ atmospheres. Given this, a logical
%Therefore the first-ever exoplanet O$_2$ detection is likely to be made on an abiotically oxidized world. A prudent 
step on the path toward biosignature detection on an Earth-like planet would be to shift the search away from Earth analogues and, instead, target small, short-period planets around low-mass stars that are predicted to possess atmospheric O$_2$ and are potentially observable with current instrumentation. Atmospheric O$_2$ detection on such worlds can then be leveraged as we expand the search to progressively more Earth-like planets.

We predict the likelihood of He, D, and O$_2$ fractionation for planets with measured masses and radii by interpolating simulated data on a three-dimensional grid of planet mass ($M_\mathrm{P}$), planet radius ($R_\mathrm{P}$), and incident stellar flux ($F_\mathrm{P}$) following \cite{Cherubim_2024}. The complete list of fractionated planet candidates is presented in Table \ref{tab:targets}. Given the uncertainty in instellation history for each planet, we included planets with $F_\mathrm{P}$ values within $\pm$ 1 dex of the interpolated $F_\mathrm{P}$ value for a given $M_\mathrm{P}$ and $R_\mathrm{P}$ in the simulated data grid.

\section{Discussion} \label{sec:discussion}

Our results build on previous efforts to model sub-Neptune atmospheric evolution by combining the effects of mass fractionation via atmospheric escape and magma ocean-atmosphere volatile exchange. Our Monte Carlo approach goes beyond models of pure H$_2$O atmospheres and Earth-like planets to provide a broad, population-level view of sub-Neptunes and super-Earths around GKM stars that start with solar composition atmospheres. We lay out testable predictions that serve to motivate future surveys, especially regarding the search for atmospheric O$_2$, and map connections between planetary interior properties like $fO_2$ and observable atmospheric signatures. While our key results are robust to many assumptions, we address some caveats below and sensitivity tests we performed to address them. Finally, we discuss some roles that photochemistry may play on our results.

\subsection{Surface Temperature} \label{sec:surface_temp}

The main controls on interior-atmosphere exchange are magma ocean volume and volatile solubility, which are determined primarily by surface temperature and mantle composition respectively. To determine surface temperature, our model assumes a dry adiabat in a troposphere defined at atmospheric pressure greater than 0.2 bar \citep{Robinson_2012}. This requires an assumption about the adiabatic index, and thus the specific heat capacity of the gas mixture in the atmosphere, which we take to be $R/0.28$ where $R$ is the specific gas constant. For reference, $R/c_{p} = 0.28$ for a N$_2$ atmosphere, 0.29 for an H$_2$ atmosphere, 0.4 for a He atmosphere, and 0.22 for a CO$_2$ atmosphere at 300 K. The surface temperature calculation is highly sensitive to the choice of adiabatic index and the only downstream calculation affected by surface temperature in our model is mantle melt fraction.

In order to evaluate the sensitivity of our main results to the surface temperature and mantle melt fraction calculations, we repeated our simulations assuming a fixed mantle melt fraction of 1.0 to represent an end member case. We found that the key result of a planetary oxidation gradient surrounding the small planet radius valley is largely unaffected, although D/H fractionation was significantly muted, rarely exceeding $\approx$ 50 times the solar value, as more H$_2$O shielding provided a larger reservoir to dilute the D/H enhancement. Our previous results represent the other extreme in which the mantle melt fraction was fixed at 0, i.e. no interior-atmosphere exchange was allowed \citep{Cherubim_2024}. 

It is possible, especially for high mean molecular weight secondary atmospheres, that convective inhibition near the planetary surface may form a radiative layer which lowers the surface temperature \citep{Selsis_2023, Innes_2023}. This would have the effect of reducing the mantle melt fraction and magma ocean volume, reducing water shielding and mantle oxidation. We assumed zero planetary albedo, $A$, for all simulations in our main results, which places an upper limit on equilibrium temperature and its influence on surface temperature. We repeated simulations assuming $A = 0.5$ and revealed that, while individual planets have different outcomes due to lower equilibrium and surface temperatures, the $(1 - A)^{(1/4)}$ equilibrium temperature dependence is insufficient to impact the planet population in our model.

\subsection{Escape Rate} \label{sec:escape_rate}

Our escape rate calculations are affected by uncertain factors including stellar flux history, escape shutoff via molecular line cooling, and the effects of ionization on atomic and molecular diffusion in the escaping wind. We adopt a simple power law parameterization to model the stellar EUV flux evolution which ignores the pre-main sequence, during which excess flux may drive additional escape. We repeated simulations around M stars using an EUV evolution model that includes the impact of stellar rotation on magnetic field evolution which captures behavior in the pre-main sequence phase \citep{Johnstone_2021}. Our key results are unaffected. This is due in large part to the radiation/recombination-limited escape calculation which limits the escape rate in spite of elevated flux during the pre-main sequence phase. Hence, the integrated EUV energy that effectively drives escape over 5 Gyr is comparable to that for simulations using the power law parameterization.

Our model ignores ionization of escaping species, except for the role of ionized hydrogen in radiation/recombination-limited escape. Ionized atomic species are known to possess significantly larger collisional cross sections due to Coulomb interactions, which will impact species diffusion in the outflow \citep{Schulik_2023}. This could potentially suppress D/H fractionation, for which the mass ratio is minimal, but is unlikely to overwrite the general predicted oxidation gradient. Our radiation/recombination-limited escape calculation accounts for dissipation of high energy radiation via recombination of protons and electrons and subsequent emission in Lyman-alpha, and captures some of the physics regarding escape suppression due to line cooling. However, our model ignores other line cooling mechanisms that may occur in the presence of the included molecular species \citep{Nakayama_2022, Yoshida_2024}. \cite{Misener_2021} propose that the core-powered mass loss mechanism may shut down when the cooling timescale equals the loss timescale, leaving behind a tenuous atmosphere. They report analytical estimates of retained atmospheric mass fractions between $10^{-4}$ and $10^{-8}$ which largely coincide with values produced in our simulations.

\subsection{Photochemistry} \label{sec:photochemistry}

%Our model ignores photochemistry, since the transition from reduced to oxidized worlds is inherently difficult for one model to capture. Photochemistry would modify our results. 
Photochemistry may affect our results in predictable ways. First, a dry CO$_2$ atmosphere would likely be susceptible to a CO-runaway state \citep[e.g.][]{Zahnle_2008, Hu_2020}, unless trace gases were available to catalytically oxidize the CO like Cl-chemistry does on Venus for example \citep{McElroy_1973}. Escape of O could also be faster on such worlds. CO$_2^+$ from photoionization of CO$_2$ is short-lived in the presence of O generated from photolysis of CO$_2$, and would produce O$_2^+$. Subsequent dissociative recombination of O$_2^+$ is the primary driver of O loss on Mars today and ion loss of O$_2^+$ could have been faster in the past \citep{Jakosky_2018}. 
% Can probably leave out depending on your definition of water world.
%Second, at water worlds, H2 may be further depleted by reactions with O and O(1D) (Yung and Demore, 1972), where atomic O would source from O2 photolysis. Similarly, CH4 would be depleted much sooner in the planet's evolution by oxidation, increasing both CO and H2. Meanwhile, CO may recycle to CO2 more efficiently, and the water involved in the reaction would source more H2 through these reactions too (e.g., Hu and Seager, 2014). 
Second, on CH$_4$ worlds, hazes are expected to form as a result of CH$_4$ photolysis, and their formation is generally controlled by an atmosphere's C:O ratio. Yet, CH$_4$ and H$_2$O are unlikely to co-exist at relatively large abundances. OH generated from water photolysis is a strong oxidant and would likely attack products of CH$_4$ photolysis before haze formation could occur. Finally, on all planets, photolysis of escaping CO$_2$ and H$_2$O would split these molecules into CO and OH respectively, which may suppress hydrodynamic escape rates due to radiative cooling effects \citep{Yoshida_2024}.

%Bower et al 2022 discuss how a surface lid could form for melt fractions $<$30\% which could trap water in the interior and maintain a carbon-rich atmosphere.

\section{Conclusions} \label{sec:conclusion}

\begin{enumerate}
\item Exchange of volatiles between an evolving magma ocean and an escaping atmosphere on sub-Neptune-mass planets creates a gradient in planetary oxidation that straddles the radius valley. Smaller planets are progressively more oxidized due to atmospheric fractionation. As primary atmospheres grow thinner, oxidized secondary atmospheres emerge due to mantle outgassing. As a result, planets that experience outgassing are able to retain replenished atmospheres at higher instellations than expected from atmospheric escape alone because volatiles can be sequestered in the mantle and shielded from escape for up to several Gyr before returning to the atmosphere.

\item A key factor in atmospheric oxidation of secondary atmospheres is the dissolution of water into the molten mantle which buffers atmospheric escape. A larger atmosphere creates a deeper, more voluminous magma ocean and higher partial pressure of water, allowing more water to dissolve into the magma ocean. This shrinks the atmosphere and therefore slows escape. As the H-dominated primary atmosphere is lost, the mantle cools and solidifies, and water outgasses into the atmosphere where it is photolyzed. H escapes and O tends to avoid escape due to low escape rates at this late evolutionary stage. Atmospheres rich in O$_2$ emerge, even if the $f$O$_2$ is buffered internally by Fe-FeO equilibria, because of the greater tendency for other, lighter volatile elements (H, He, C) to escape relative to O. %O returns to the mantle depending on the availability of electrons, which is largely driven by the iron abundance and oxidation state. The result is an oxidized mantle and atmosphere.

\item We present robust predictions of a broad population of planets with O$_2$-dominated atmospheres, for the first time starting from solar composition sub-Neptunes. O$_2$ worlds ($x_\mathrm{O2} > 0.5$) comprise the majority ($\approx 70\%$) of all simulated planets with secondary atmospheres. Most exist on close-in orbits around low mass stars, making them amenable to atmospheric characterization through transit spectroscopy. Current pursuits of the first-ever exoplanet atmospheric O$_2$ detection are overwhelmingly focused on Earth analogues, which are not amenable to atmospheric characterization. In addition, such efforts are motivated by a much more complex mechanism for O$_2$ production, i.e., global oxygenic biospheres, compared to the well understood abiotic O$_2$ buildup mechanism of H$_2$O photolysis and H escape. As a natural step on the exciting path toward biosignature discovery, we call on the community to shift the focus of O$_2$ searches to close-in, abiotic O$_2$ worlds.

\item Helium worlds remain a robust population of radius valley planets predicted by our model, and are still observationally unconfirmed, motivating future observing campaigns. This newly emerging planetary class presents a novel climate regime and may be the most observable sub-Neptune class with processed, secondary atmospheres. Helium worlds may in fact be unique as the only sub-Neptunes with relatively high scale heights that can be oxidized. Our simulations reveal that weakly helium-enriched planets are methane-rich, while helium-dominated planets possess molecular oxygen, carbon monoxide, and carbon dioxide. Deuterium-enriched worlds with high D/H ratios are also predicted by our model and deuterated species may be observable via atmospheric spectroscopy.
\end{enumerate}

\software{{\tt SciPy} \citep{scipy}, {\tt Atmodeller} \citep{Bower_2025},
{\tt IsoFATE} \citep{Cherubim_2024}
          }

%\begin{acknowledgments}
\section*{Acknowledgments}

% This research has made use of the NASA Exoplanet Archive, which is operated by the California Institute of Technology, under contract with the National Aeronautics and Space Administration under the Exoplanet Exploration Program. The data can be accessed via \dataset[NASA Exoplanet Archive]{\doi{10.26133/NEA13}}.
This research has made use of the NASA Exoplanet Archive \citep{PSCompPars}\footnote{Accessed on 2024-10-07 at 10:14, returning 287 rows.)} This dataset or service is made available by the NASA Exoplanet Science Institute at IPAC, which is operated by the California Institute of Technology under contract with the National Aeronautics and Space Administration.

RW acknowledges funding from the Leverhulme Center for Life in the Universe, Joint Collaborations Research Project Grant G119167, LBAG/312.

DJB and PAS were supported by the Swiss State Secretariat for Education, Research and Innovation (SERI) under contract No. MB22.00033, a SERI-funded ERC Starting grant “2ATMO”. PAS also thanks the Swiss National Science Foundation (SNSF) through an Eccellenza Professorship (203668).

DJA is funded by NASA through the NASA Hubble Fellowship Program Grant HST-HF2-51523.001-A awarded by the Space Telescope Science Institute, which is operated by the Association of Universities for Research in Astronomy, Inc., for NASA, under contract NAS5‐26555. 

%\end{acknowledgments}

\bibliographystyle{aasjournal}
\bibliography{refs.bib}

\end{document}